\documentclass[sn-mathphys-num]{sn-jnl}% Basic Springer Nature Reference Style/Chemistry Reference Style
\usepackage{graphicx}%
\usepackage{multirow}%
\usepackage{amsmath,amssymb,amsfonts}%
\usepackage{amsthm}%
\usepackage{mathrsfs}%
\usepackage[title]{appendix}%
\usepackage{xcolor}%
\usepackage{textcomp}%
\usepackage{manyfoot}%
\usepackage{booktabs}%
\let\cline\cmidrule

\usepackage{listings}%
\usepackage{natbib}%

\usepackage{graphicx}
\usepackage{amssymb}
\usepackage{xspace}
\usepackage{float}
\usepackage{xcolor}
\usepackage{multicol}
\usepackage[labelformat=empty]{subcaption}
\usepackage[ruled,vlined,linesnumbered]{algorithm2e}

\usepackage{amsmath}

\usepackage{todonotes}
\usepackage{comment}

\newcommand{\BWT}{\ensuremath{\text{BWT}}\xspace}
\newcommand{\SA}{\ensuremath{\text{SA}}\xspace}
\newcommand{\LCP}{\ensuremath{\text{LCP}}\xspace}
\newcommand{\LCS}{\ensuremath{\text{LCS}}\xspace}
\newcommand{\lcpeq}[2]{{\text{lcp$(#1,#2)$}}}

\newcommand{\KL}{\ensuremath{\text{KL}}\xspace}
\newcommand{\DA}{\ensuremath{\text{DA}}\xspace}
\newcommand{\BWSD}{\ensuremath{\text{BWSD}}\xspace}
\newcommand{\Node}{\ensuremath{{\it Node}}\xspace}
\newcommand{\SL}{\ensuremath{\text{CL}}\xspace}

\renewcommand{\S}{\ensuremath{\mathcal{S}}\xspace}

\newcommand{\gcBB}{\ensuremath{\texttt{gcBB}}\xspace}
\newcommand{\mgcBB}{\ensuremath{\texttt{multi-gcBB}}\xspace}
\newcommand{\DE}{\ensuremath{\text{$D_E$}}\xspace}
\newcommand{\DM}{\ensuremath{\text{$D_M$}}\xspace}

\newcommand{\maxlcp}{\mathsf{maxlcp}}
\newcommand{\red}{\textcolor{red}}
\newcommand{\blue}{\textcolor{blue}}

\definecolor{orange}{rgb}{1.0,0.5,0.0}

\newtheorem{definition}{Definition}%

\begin{document}

\title[Article Title]{Comparative genomics with succinct colored {de~Bruijn} graphs}

%%=============================================================%%
%% GivenName    -> \fnm{Joergen W.}
%% Particle     -> \spfx{van der} -> surname prefix
%% FamilyName   -> \sur{Ploeg}
%% Suffix       -> \sfx{IV}

\author*[1]{\fnm{Lucas P.} \sur{Ramos}}\email{lucaspr98@gmail.com}
\equalcont{These authors contributed equally to this work.}

\author[2]{\fnm{Felipe A.} \sur{Louza}}\email{louza@ufu.br}
\equalcont{These authors contributed equally to this work.}

\author[1]{\fnm{Guilherme P.} \sur{Telles}}\email{gpt@ic.unicamp.br}
\equalcont{These authors contributed equally to this work.}

\affil*[1]{\orgdiv{Instituto de Computação}, \orgname{UNICAMP}, \orgaddress{\city{Campinas}, \state{SP}, \country{Brazil}}}

\affil[2]{\orgdiv{Faculdade de Engenharia El\'etrica}, \orgname{UFU}, \orgaddress{\city{Uberl\^andia},  \state{MG}, \country{Brazil}}}


\abstract{
DNA technologies have evolved significantly in the past years enabling the sequencing of a large number of genomes in a short time. Nevertheless, the underlying computational problem is hard, and many technical factors and limitations complicate obtaining the complete sequence of a genome.  Many genomes are left in a draft state,  in which each chromosome is represented by a set of sequences with partial information on their relative order. Recently, some approaches have been proposed to compare draft genomes by comparing paths in de Bruijn graphs, which are constructed by many practical genome assemblers. In this article we introduce \gcBB, a method for comparing genomes represented as succinct colored de Bruijn graphs directly, without resorting to sequence alignments, by means of the entropy and expectation measures based on the Burrows-Wheeler Similarity Distribution. We also introduce an improved version of \gcBB, called \mgcBB, that improves the time performance considerably through the selection of different data structures. We have compared phylogenies of genomes obtained by other methods to those obtained with \gcBB, achieving promising results.
}




\maketitle

\section{Introduction}
Genome assembly is the task of reconstructing the sequence of
nucleotides in molecules of DNA that follows the DNA sequencing process, in which a large amount of
short strings representing fragments of consecutive nucleotides in the target DNA molecule (reads) is first obtained.
The reads cover each DNA nucleotide many times, varying across
sequencing projects and
may be as high as 200 times per nucleotide.  The reads
must then be assembled based on the overlaps among them.  This is a
hard computational problem, further complicated by the huge number of
reads that may be obtained with the current DNA sequencing
technologies, by the presence of repetitions in the target DNA, by
sequencing errors and by other sources of ambiguities and technical
limitations.  Completely assembling a genome also requires intensive wet-lab work, then
many genomes are left in a draft state after sequencing, that is, instead of a single string for each chromosome there is a set of strings (contigs) that may include information on their relative order (scaffolds)~\cite{2019-rice}.

Several approaches have been proposed to assemble genomes.  Some of
the most used are based on different assembly graphs, such as
overlap graphs~\cite{rizzi2019overlap}, de Bruijn
graphs~\cite{de1946combinatorial}, string graphs~\cite{SGA} and
repeat graphs~\cite{kolmogorov2020metaflye}.  These assembly graphs
can also be useful in gene discovery, structural variation
analysis, hybrid assembly and other
applications~\cite{scheibye2009sequence}.

Comparative genomics aims at identifying similar and dissimilar
regions among genomes~\cite{Makinen2015}.  Through comparison it is possible, for
instance, to identify conserved regions across species that may be
related to cellular processes, to identify regions involved in
mutational events, to build phylogenies based on similarity among sequences, etc.

In Bioinformatics, the most widely employed means of calculating
similarity between biological sequences is through alignments~\cite{KimJY20}.
When used with evolution models for amino acids and nucleotides,
alignments provide a similarity measure that reflects the
evolutionary distance between molecules and is supported by
statistics that are readily understood by practitioners~\cite{Rosenberg05}. Alignments
also provide a natural layout for visualizations and visual data
exploration~\cite{ZhuNCWSL13}.

On the other hand, the quadratic computational cost of the
algorithms to calculate alignments between strings ($O(nm)$ time for
two strings of lengths $n$ and $m$) coupled with the huge amount of
data currently available in public and private repositories push
the need for faster alternatives such as heuristics, parallel
algorithms and alternative distance measures, including
alignment-free strategies~\cite{Makinen2015}.
When it comes to the comparison of multiple genomes, producing a
multiple sequence alignment is even harder computationally, and the use of heuristics is widespread.

Similarity measures for strings in general, not only DNA or protein,
may be computed in many ways, for example as distance among vector
representation of strings, as statistics calculated on groups of symbol co-occurrences, as edit distance, as alignment score, as substring tiling  and
others~\cite{setubal1997introduction,baeza1999modern}.
Compression-based measures also exist, such as the NCD~\cite{ncd},
whose idea originates in the works on minimum algorithmic descriptions of strings.
Similarity measures based on the Burrows-Wheeler transform
(BWT)~\cite{burrows1994block}, as the eBWT-based
distances~\cite{MANTACI2007298,MantaciRS08} and the Burrows-Wheeler
Similarity Distribution (BWSD)~\cite{YANG2010742}, are particularly
attractive because the BWT provides a self-index~\cite{Navarro2016}
that can be computed in linear time on the string length.

Many genome assemblers are based on the de Bruijn graph ({\it
  e.g.}~\cite{BWA,Bowtie2,SGA}), that may be stored succinctly using
the BOSS representation~\cite{wabi/BoweOSS12}.  The BOSS
representation is based on the BWT and enables the assembly of
larger sets of reads.  Colors may be added to the edges of a de
Bruijn graph, allowing the representation of a set of strings from
distinct genomes on colored de Bruijn graphs.  Recent approaches
have been proposed for the comparison of genomes by the extraction
of paths from their colored de Bruijn
graphs~\cite{lyman2017whole,polevikov2019synteny}.

In this paper we introduce \gcBB\footnote{A preliminary version of this work appeared in~\cite{RamosLT22}.},
a space-efficient algorithm for comparing genomes using succint de Bruijn graphs and BWSD.
Given a set of genomes, each one represented by a set of  unmounted raw reads, a colored de Bruijn graph in the BOSS
representation is built for all
the genomes and BWSD-based
measures are evaluated to assess the similarity between all genomes in the set.
Our algorithm computes the
colored de Bruijn graph for all genomes only once, avoiding a pairwise construction, and the
BWSD-based measures are computed using compressed data structures as proposed in~\cite{tcs/LouzaTGZ19}.
Our method showed promising results in experiments that compared the
phylogenies for genomes of 12 Drosophila species built with $\gcBB$
and with the methods by Lyman \textit{et
  al.}~{\cite{lyman2017whole}} and by Polevikov and
Kolmogorov~{\cite{polevikov2019synteny}}.

\section{Definitions and notation}

A string is the juxtaposition of symbols from an ordered alphabet
$\Sigma$ of size $\sigma$.  Let $S$ be a string of length $n$. We
index its symbols from $1$ to $n$.  A substring of $S$ is $S[i,j] =
S[i] \dots S[j]$ with $1 \leq i \leq j \leq n$.  Whenever $i >j$
then $S[i,j]$ denotes the empty string. Any substring $S[1,i]$ is
referred to as a prefix of $S$ and $S[i,n]$ is referred to as a
suffix of $S$.  The concatenation of string and symbols will be
denoted by juxtaposition.

For clearer definitions we assume that the last symbol of a string
$S$ is the special end-marker symbol $\$$, that does not occur elsewhere
in $S$ and is the smallest symbol in $\Sigma$.  This way, all
suffixes of $S$ are distinct.

The suffix array~\cite{manber1993suffixarray} of a string $S$ of
length $n$ is the array $\SA_S$ containing the permutation of
$\{1,\ldots,n\}$ that gives the suffixes of $S$ in lexicographic
order, that is, $S[\SA_S[1],n] < S[\SA_S[2],n] < \ldots < S[\SA_S[n],n]$.

By \lcpeq{S_1}{S_2} we denote the length of the longest common
prefix of strings $S_1$ and $S_2$.  The \LCP array for a string $S$
of length $n$ is the array of integers containing the lcp of
consecutive suffixes in $\SA_S$. Formally,
$\LCP_S[i]=\lcpeq{S[\SA_S[i],n]}{S[\SA_S[i-1],n]}$ for $1 < i \leq
n$ and $\LCP_S[1] = 0$.  Figure~\ref{fig:saBWT} shows the suffix
array and the \LCP array for the string $S=\texttt{abracadabra\$}$.

\begin{figure}[t]

\begin{center}
\small
\begin{tabular}{c|c|c|c|l|}
\cline{2-5}
$i$ & $\SA_S$ &  $\LCP_S$ & $\BWT_S$    & $S[\SA_S[i],n]$ \\
\cline{2-5}
1   & 12      & 0   & a          & \$                 \\
2   & 11      & 0   & r          & a\$                \\
3   & 8       & 1   & d          & abra\$             \\
4   & 1       & 4   & \$         & abracadabra\$     \\
5   & 4       & 1   & r          & acadabra\$        \\
6   & 6       & 1   & c          & adabra\$           \\
7   & 9       & 0   & a          & bra\$              \\
8   & 2       & 3   & a          & bracadabra\$      \\
9   & 5       & 0   & a          & cadabra\$         \\
10  & 7       & 0   & a          & dabra\$           \\
11  & 10      & 0   & b          & ra\$               \\
12  & 3       & 2   & b          & racadabra\$       \\
\cline{2-5}
\end{tabular}%
\end{center}
\caption{Suffix array, LCP array and BWT for string
  $S=\text{abracadabra\$}$.  The last column shows the suffixes of
  $S$ in the order provided by the suffix array.}
\label{fig:saBWT}
\end{figure}

Let $\S=\{S_1, S_2, \ldots, S_d\}$ be a collection of $d$ strings of
lengths $n_1,n_2,\ldots, n_d$.  We define the concatenation of all
strings in $\S$ as $S^{cat}=S_1[1,n_1-1]\$_1 S_2[1,n_2-1] \$_2
\cdots S_d[1, n_d-1] \$_d$, that is, the end-marker symbol ${\$}$ of
each string is replaced by a separator symbol $\$_i$ such that $\$_i
< \$_j$ if $i < j$ and every $\$_i<\$$.  The length of the
concatenated string $S^{cat}$ is $N = \sum_{i=1}^d n_i$.  The suffix
and $\LCP$ arrays for a collection $\S$ correspond to the arrays
$\SA_{\S}[1,N]$ and $\LCP_{\S}[1,N]$ computed for $S^{cat}$.

We define the context of a suffix $S^{cat}[i, N]$ as the substring
$S^{cat}[i,j]$ such that $S^{cat}[j]$ is the leftmost occurrence of
some $\$_k$ in $S^{cat}[i, N]$.  The document array is an array of
integers $\DA_{\S}$ of length $n$ that stores to which string each
suffix in $\SA_{\S}$ belongs. Formally, $\DA_{\S}[i] = j$ if the
context of $S^{cat}[\SA_{\S}[i],N]$ ends with $\$_j$.  The suffix,
\LCP and document arrays for a collection $\S$ can be computed in
linear time using constant workspace~{\cite{tcs/LouzaGT17}}.

When clear from the context, we drop the subscripts in $\SA_S$, $\LCP_S$,
$\SA_{\S}$ and $\DA_{\S}$.
Without loss of generality, we assume that the alphabet $\Sigma$ of a string is $\{1,\ldots,\sigma\}$ or has been implicitly mapped onto $\{1,\ldots,\sigma\}$.

\subsection{BWT and BWSD}

The Burrows-Wheeler Transform (BWT)~\cite{burrows1994block} of a
string $S$ is a reversible transformation that permutes its symbols
such that the resulting string, denoted by $\BWT_S$ or simply by
\BWT when the context is clear, often allows better compression
because equal symbols tend to be clustered.  The BWT is the core of many indexing structures for
text~\cite{ohlebusch2013bioinformatics,Makinen2015,Navarro2016}.

For $0\leq i < n$, the $i$-th circular rotation (or conjugate or
simply rotation) of a string $S$ is the string $S[i+1,n]S[1,i]$.
As $S[n]=\$$, its rotations are distinct.
The \BWT is the last column of a matrix $\mathcal{M}$ having the
sorted rotations of $S$ as rows.  In $\mathcal{M}$, the first column
is called $F$ and the last column is called $L$.  Since $S[n]=\$$,
sorting the rotations of $S$ is equivalent to sorting the suffixes
of $S$ and then the \BWT may be defined in terms of the suffix array
as $\BWT[i] = S[\SA[i] - 1]$ if $\SA[i] \neq 1$ or $\BWT[i]=\$$
otherwise.  The \BWT for $S=\texttt{abracadabra\$}$ is shown in
Figure~\ref{fig:saBWT}.  The BWT for a collection of strings $\S$
may be obtained from the SA of $S^{cat}$ as well~\cite{CenzatoL22}.

The \BWT of strings $S_1$ and $S_2$ may be used to compute
similarity measures between them based on the observation
that the amount of symbols of $S_1$ intermixed with symbols of $S_2$
in the \BWT of $S_1S_2$ is related to
the amount of substrings shared by $S_1$ and
$S_2$~\cite{MANTACI2007298}.

The Burrows-Wheeler similarity distribution
(\BWSD)~\cite{YANG2010742} between $S_1$ and $S_2$, denoted by
$\BWSD(S_1,S_2)$, is a probability mass function defined as follows.
Given the \BWT of $\S = \{S_1, S_2\}$, we define a bitvector
$\alpha$ of size $n_1+n_2$ such that $\alpha[p] = 0$ if $\BWT[p] =
\$_2$ or $\BWT[p] \in S_1$ and $\alpha[p]=1$ if $\BWT[p] = \$_1$ or
$\BWT[p] \in S_2$.  The bitvector $\alpha$ can be represented as a
sequence of runs
\begin{equation*}
r=0^{k_1}1^{k_2}0^{k_3}1^{k_4} \ldots 0^{k_m}1^{k_{m+1}}
\end{equation*}
where $i^{k_j}$ means that $i$ repeats $k_j$ times and only $k_1$ and
$k_{m+1}$ may be zero. The largest possible value for $k_j$ is $\max\{n_1,n_2\}$. We denote $\max\{k_j | i^{k_j} \in r\}$ by $k_{\max}$.

% kmax está mal... Não é n_1 ou n_2.

Let $t_k$ be the number of occurrences of an exponent $k$ in $r$. Let
$s=t_1+t_2+\ldots+t_{k_j}+\ldots+t_{k_{\max}}$. The $\BWSD(S_1,S_2)$
is the probability mass function
\begin{equation*}
P\{k_j=k\}=t_k/s \mbox{ for } k=1,2,\ldots,k_{\max}
\end{equation*}

\noindent
Two distance measures were defined on the BWSD of $S_1$ and $S_2$~\cite{YANG2010742}.

\begin{definition}
The expectation distance is $D_M(S_1,S_2)=E({k_j})-1$, where $E(k_j)$ is the expectation of $\BWSD(S_1,S_2)$.
\end{definition}

\begin{definition}
The entropy distance is $D_E(S_1,S_2)=\mathcal{H}(\BWSD(S_1,S_2))$, where
$\mathcal{H}(\BWSD(S_1,S_2)) = -\sum_{k\geq 1, t_k \neq 0}(t_k/s)\log_2(t_k/s)$ is the Shannon entropy of $\BWSD(S_1,S_2)$.
\end{definition}

% Essa def sempre me incomodou porque a equação e o texto estão dizendo a mesma coisa. Coloquei um that is para ver.

Note that if $S_1$ and $S_2$ are equal then $k_{\max}=1$, $P\{k_j=1\}=
\frac{n_1+n_2}{n_1+n_2} = 1$, $D_M(S_1,S_2)=0$ and $D_E(S_1,S_2)=0$.
Also, if $\alpha$ for $\BWT(S_1,S_2)$ is equal to $\alpha$ for $\BWT(S_2,S_1)$,
then both have the same BWSD, and $D_E(S_1,S_2)=D_E(S_2,S_1)$ and $D_M(S_1,S_2)=D_M(S_2,S_1)$.

\subsection{Succinct de Bruijn graphs}\label{sec:deBruijnBOSS}

Let $\S=\{S_1, S_2, \ldots, S_d\}$ be a collection of strings (reads
of a genome).  Assume that $S$ is modified by concatenating $k$
symbols $\$$ at the beginning of each string in $\S$.  We will refer
to a string of length $k$ as a $k$-mer.

\begin{figure}
    \centering
    \includegraphics[width=0.65\textwidth]{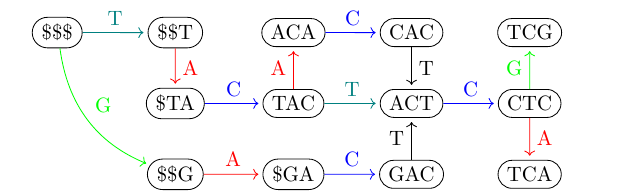}
    \caption{
    De Bruijn graph for $\S=\{{\tt \$\$\$TACACT, \$\$\$TACTCA, \$\$\$GACTCG}\}$ and $k=3$.
    %Starting at node {\tt \$\$\$} and walking through the edges labelled with {\tt T,A,C,A,C,T} successively, we obtain the first string of the collection.
    }
    \label{fig:dbg(a)}
\end{figure}

A de Bruijn graph (of order $k$) for $\S$ has one vertex for each distinct $k$-mer in a string of $S$.
We say that the $k$-mer related to a vertex $u$ is its vertex label, denoted
by $\overrightarrow{u}$.
There is an edge labeled $v[k]$ from vertex $u$ to
vertex $v$ if the substring
$u[1]u[2]{\ldots}u[k]v[k]$ occurs in a string of $\S$.
For example, given $\S=\{{\tt \$\$\$TACACT, \$\$\$TACTCA, \$\$\$GACTCG}\}$ and $k=3$, Figure~{\ref{fig:dbg(a)}} illustrates the de Bruijn graph of order $k$ for $\S$.

Notice that an edge from $u$ to $v$ corresponds to the existence
of an overlap of length $k-1$ between the suffix of
$\overrightarrow{u}$ and the prefix of $\overrightarrow{v}$ and
also that
the concatenation of edge labels along a path of length $k$ that
arrives at a vertex $v$ whose label does not have a $\$$ will be
$\overrightarrow{v}$.
For example, in Figure~{\ref{fig:dbg(a)}}, starting at node {\tt \$\$\$} and traversing the edges labelled {\tt T,A,C,A,C,T} successively, we obtain the first string in the collection.

BOSS~\cite{wabi/BoweOSS12} is a succinct representation of the de
Bruijn graph that enables efficient navigation across vertices and
edges.  Let $n$ and $m$ be respectively the number of vertices and
edges of a de Bruijn graph $G$.  Assume that the vertices $v_1, v_2,
\ldots, v_n$ in $G$ are sorted according to the co-lexicographic
order of their labels, {\it i.e.}, the lexicographic order of the
reverse of their labels,
$\overleftarrow{v_i}=\overrightarrow{v_i}[k] \ldots
\overrightarrow{v_i}[1]$ for each vertex $v_i$.

We define $\Node$ as a conceptual matrix containing the
co-lexicographically sorted set
with the
distinct $k$-mers in $\S$ and with,
for each vertex that has $t>1$ outgoing edges with distinct labels,
$t-1$ additional copies of that label. 
Let $m$ be the number of rows in $\Node$.

For each vertex $v_i$, we define $W_i$ as the sequence of symbols of
the outgoing edges of $v_i$ in lexicographic order. If $v_i$ has no
outgoing edges then $W_i=\$$.

The BOSS representation is composed by the following components:

\begin{enumerate}
\item The string $W[1,m] = W_1W_2\dots W_n$. Observe that $|W| =
  |\Node|$ and $\Node[i]$ denotes the vertex from which $W[i]$ leaves.

\item The bitvector $W^-[1,m]$ such that $W^-[i]=0$ if there exists
  $j<i$ such that $W[j] = W[i]$ and the suffixes of length $k-1$ of
  $\Node[j]$ and of $\Node[i]$ are identical, or $W^-[i]=1$ otherwise.

\item The bitvector $last[1,m]$ such that $last[i]=1$ if $i=n$ or
  $\Node[i]$ is different from $\Node[i+1]$, or $last[i]=0$ otherwise.

\item The counter array $C[1,\sigma]$ such that $C[c]$ stores the
  number of symbols smaller than $c$ in the last column of the
  conceptual matrix $\Node$.

\end{enumerate}

\begin{figure}
    \centering
    %\begin{subfigure}[c]{0.55\textwidth}
    \includegraphics[width=0.6\textwidth]{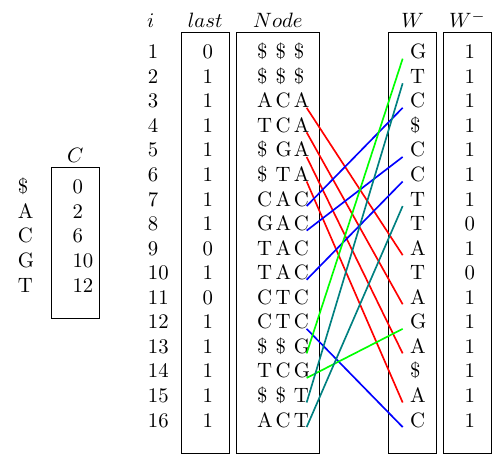}
    % \caption{(a)}
    %\end{subfigure}
    %\hfill
    %\begin{subfigure}[c]{0.65\textwidth}
    %\includegraphics[width=\textwidth]{images/boss-dbg.pdf}
    %\caption{(b)}
    %\end{subfigure}
    \caption{(a) BOSS representation for $\S=\{{\tt \$\$\$TACACT,
        \$\$\$TACTCA, \$\$\$GACTCG}\}$ augmented with the \Node
      matrix and with edges of the de Bruijn graph; edges with the
      same color have the same symbol.}
      %and edges colored black
      %represent edges where $W^-[i] = 0$, that is, there is already
      %one or more edge $W[j]$, such that $j<i$, mapped to the same
      %vertex, to which $W[i]$ is mapped.  \red{Não tem aresta colorida de black}}
    \label{fig:boss1}
\end{figure}

For example, the succinct representation of the de Bruijn graph for
$\S=\{{\tt \$\$\$TACACT, \$\$\$TACTCA, \$\$\$GACTCG}\}$ is
illustrated in Figure~{\ref{fig:boss1}} augmented with the $\Node$
matrix and with edges of the de Bruijn graph to ease the
understanding.

For DNA sequences, the alphabet is $\Sigma = \{{\tt A,T,C,G,N,\$}\}$
with size $\sigma = 6$.  Storing the string $W$ requires $m \lceil
\log_2 \sigma \rceil = 3m$ bits, the bitvectors $W^-$ and $last$
require $2m$ bits and the counter array $C$ requires $\sigma \log_2 m
= 6 \log_2 m$ bits. Therefore, the overall space to store
the BOSS structure is $5m + 6 \log_2 m$ bits. 

Egidi {\it et al.}~\cite{egidi2019external} proposed an algorithm
called eGap for computing the multi-string BWT and the LCP array in
external memory.  As an application the authors showed how to
compute the BOSS representation with a sequential scan over the \BWT
and the \LCP array built for collection $\S$ with all strings
reversed in $O(N)$ time.

The \emph{colored de Bruijn graph}~{\cite{iqbal2012novo}}
generalizes the formulation of a de Bruijn graph for a set
$\S=\{\S_1,\ldots,\S_g\}$ of $g$ string collections.
In the colored de Bruijn graph the set of vertices includes all
strings in $\S$ and there may be parallel edges with the same label,
but with distinct colors.
Formally, the colored de Bruijn graph (of order $k$) for
$\S=\{\S_1,\ldots,\S_g\}$ has one vertex for each distinct $k$-mer
in a string of $\S$.
There is an edge labeled $v[k]$ and colored $i$ from vertex $u$ to
vertex $v$ if the substring $u[1]u[2]{\ldots}u[k]v[k]$ occurs in a
string of $\S_i$.

\begin{figure}
  \centering
  \begin{subfigure}{0.5\textwidth}
    \includegraphics[width=\linewidth]{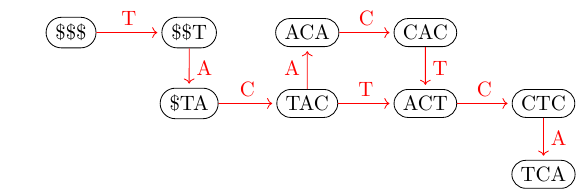}
    \caption{(a)} \label{fig:1a}
  \end{subfigure}%
  \begin{subfigure}{0.5\textwidth}
    \includegraphics[width=\linewidth]{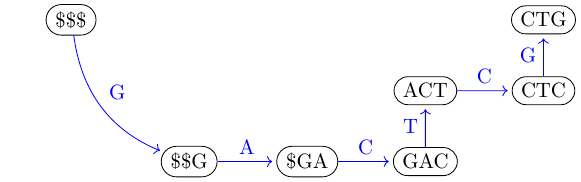}
    \caption{(b)} \label{fig:1b}
  \end{subfigure}%
  \\
  \begin{subfigure}{0.55\textwidth}
    \includegraphics[width=\linewidth]{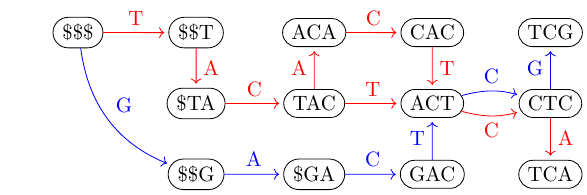}
    \caption{(c)} \label{fig:1c}
  \end{subfigure}
  \caption{(a) De Bruijn graph for $\S_1=\{{\tt\$\$\$TACACT,
      \$\$\$TACTCA}\}$. (b) De Bruijn graph for
    $\S_2=\{{\tt\$\$\$GACTCG}\}$. (c) Colored de Bruijn graph for
    $\{\S_1, \S_2\}$, where red edges are from $\S_1$ and blue edges
    are from $\S_2$.  We remark that only the graph for $\{\S_1,
    \S_2\}$ has colored edges, $\S_1$ and $\S_2$ edges are colored
    for example purposes.}
\label{fig:colored1}
\end{figure}

For example, for collections $\S_1=\{{\tt \$\$\$TACACT,
  \$\$\$TACTCA}\}$ and $\S_2=\{{\tt\$\$\$GACTCG}\}$ and $k=3$,
Figure~{\ref{fig:colored1}} shows the de Bruijn graphs for $\S_1$
and for $\S$, and the colored de Bruijn graph for $\{\S_1, \S_2\}$.

When $|\S|=2$, a bitvector ${\tt colors}[1,m]$ is added to obtain the BOSS representation of a colored de Bruijn graph,   indicating from which genome each edge came from.  We deal with the   case $|\S|>2$ in Section~{\ref{s:mgcbb}}.

The number of reads that include a given nucleotide in the target
DNA is refered to as coverage, and may be as high as 200 with
current sequencing technologies. The coverage is used by genome
assemblers to solve ambiguities during the reconstruction.  Coverage
is also directly related to the existence of repeated regions in the
genome and to sequencing errors.

This information is handled using a {\tt coverage}.
The ${\tt coverage}[1,m]$ array stores the number of times a $(k+1)$-mer represented by an edge occurs in its genome and can be computed from the $\LCP$ array during BOSS construction.

\section{gcBB}

The input for our algorithm, called \gcBB (genome comparison using
BOSS and BWSD), is a set of genomes (each one as a
FASTQ file of reads) and a value for $k$.  For each
pair of genomes \gcBB constructs the colored BOSS and computes the
BWSD, producing two distance matrices, \DM and \DE, with the
expectation and entropy distances among all pairs of genomes in the
input set.

The intuition is that intermixed edges in the colored BOSS are
related to shared nodes in their graphs and to similarities in the
genomes.

The pseudo-code for \gcBB is shown in Algorithm~\ref{alg:gcBB}.  \gcBB
has three phases, indicated in the pseudo-code and detailed below.

\begin{algorithm}[h]
\SetAlgoLined
\KwIn{Collection $\S=\{\S_1,\ldots,\S_g\}$ and $k$-mer length}
\KwOut{Matrices $\DM$ and $\DE$ of double precision numbers}
\tcp{Phase 1}
\For{each genome $\S_i$ in \S}{
    ${\tt eGap}(\S_i)$\tcp*{compute $\LCP_i$, $\BWT_i$ and $\SL_i$}
}
\For{each pair of genomes $\{\S_i, \S_j\}$ in \S}{
    ${\tt eGap}(\S_i, \S_j)$\tcp*{merge \LCP, \BWT and \SL and generate \DA}
}
$\text{double } \DM[1..g][1..g]=0.0$\;
$\text{double } \DE[1..g][1..g]=0.0$\;
\For{each pair of genomes $\{\S_i, \S_j\}$ in \S}{
    \tcp{Phase 2}
    $\text{bitvector ${\tt colors}$ initialized with 0} $\;
    ${\tt colors} = {\tt colored\_BOSS\_construction}(\LCP_{i,j}, \BWT_{i,j}, \SL_{i,j}, \DA_{i,j}, k)$\;
    \tcp{Phase 3}
    $\text{double {\tt expectation}}=0.0$\;
    $\text{double {\tt entropy}}=0.0$\;
    $\{{\tt expectation, entropy}\} = {\tt BWSD\_computation}({\tt colors}, \LCP_{i,j}, \SL_{i,j}, k)$\;
    $\DM[i][j]= {\tt expectation}$\;
    $\DE[i][j]= {\tt entropy}$\;
}
\Return{\DM, \DE}\;
\caption{gcBB}
\label{alg:gcBB}
\end{algorithm}

\paragraph{Phase 1:}
First, \gcBB constructs the \BWT and the \LCP array for each genome
$\S_1, \S_2, \dots, \S_g$ in external memory using
eGap~\cite{egidi2019external}.  It also computes an auxiliary array
with the length of each context,
called \SL.  We remark that one could use any other tool to
construct these data structures, for
example~\cite{BonizzoniVPPR19,LouzaTGPR20,PrezzaR21}.

For each pair of genomes $\S_i$ and $\S_j$, the corresponding arrays
are merged with eGap while the document array $\DA_{i,j}$ is
computed.  Note that $\DA_{i,j}$ can be stored in a bitvector, since
we merge only pairs of genomes.  The resulting arrays are written to
external memory.

For genomes $\S_1=\{{\tt TACTCA, TACACT}\}$ and $\S_2=\{{\tt
  GACTCG}\}$, Figure~\ref{fig:s1s2eGap} shows the output of eGap for
each genome and the resulting merge.

\begin{figure}[t]
\resizebox{0.875\textwidth}{!}
{

  \begin{minipage}{.5\textwidth}
    \begin{subfigure}[b]{\linewidth}
    \centering
      \begin{tabular}{r|c|c|c|l|}
    \cline{2-5}
      $i$ & $\BWT_{2}$ & $\LCP_{1}$ & $\SL_{1}$ & context  \\
    \cline{2-5}
      1 & {\tt T}      & 0 & 1 & {\tt $\$_1$}           \\
      2 & {\tt T}      & 0 & 1 & {\tt $\$_2$}          \\
      3 & {\tt C}      & 0 & 5 & {\tt ACAT$\$_2$}     \\
      4 & {\tt $\$_2$} & 2 & 7 & {\tt ACTCAT$\$_1$}        \\
      5 & {\tt C}      & 1 & 3 & {\tt AT$\$_1$}             \\
      6 & {\tt C}      & 2 & 3 & {\tt AT$\$_2$}         \\
      7 & {\tt T}      & 0 & 6 & {\tt CACAT$\$_2$}        \\
     8 & {\tt A}      & 2 & 4 & {\tt CAT$\$_2$}        \\
     9 & {\tt T}      & 3 & 4 & {\tt CAT$\$_1$} \\
     10 & {\tt A}      & 1 & 6 & {\tt CTCAT$\$_1$}      \\
     11 & {\tt A}      & 0 & 2 & {\tt T$\$_1$}         \\
     12 & {\tt A}      & 1 & 2 & {\tt T$\$_2$}          \\
     13 & {\tt $\$_1$} & 1 & 7 & {\tt TCACAT$\$_2$}       \\
     14 & {\tt C}      & 3 & 5 & {\tt TCAT$\$_1$}         \\
    \cline{2-5}
    \end{tabular}
    \caption{(a)}
    \label{fig:s1eGap}
  \end{subfigure}%
  \\
  %\hspace*{\fill}
  \begin{subfigure}[b]{\linewidth}
  \centering
    \begin{tabular}{r|c|c|c|l|}
    \cline{2-5}
      $i$ & $\BWT_{2}$ & $\LCP_{2}$ & $\SL_{2}$ & context  \\
    \cline{2-5}
      1 & {\tt G}      & 0 & 1 & {\tt $\$_1$}     \\
      2 & {\tt C}      & 0 & 3 & {\tt AG$\$_1$}    \\
      3 & {\tt T}      & 0 & 4 & {\tt CAG$\$_1$}        \\
      4 & {\tt G}      & 1 & 6 & {\tt CTCAG$\$_1$}  \\
      5 & {\tt A}      & 0 & 2 & {\tt G$\$_1$}       \\
      6 & {\tt $\$_1$} & 1 & 7& {\tt GCTCAG$\$_1$} \\
      7 & {\tt C}      & 0 & 5 & {\tt TCAG$\$_1$}   \\
    \cline{2-5}
    \end{tabular}
    \caption{(b)} \label{fig:s2eGap}
  \end{subfigure}%
  \end{minipage}
  ~~
  \begin{minipage}{.5\textwidth}
  \begin{subfigure}[t]{\linewidth}
  \centering
    \begin{tabular}{r|c|c|c|c|l|}
        \cline{2-6}
          $i$ & $\BWT_{1,2}$ & $\LCP_{1,2}$ & $\SL_{1,2}$ & $\DA_{1,2}$  & context \\
        \cline{2-6}
          1  & {\tt T}             & 0 & 1  & 0 & {\tt $\$_1$}  \\
          2  & {\tt T}             & 0 & 1 & 0  & {\tt $\$_2$}  \\
          3  & {\tt G}             & 0 & 1 & 1  & {\tt $\$_3$} \\
          4  & {\tt C}             & 0 & 5 & 0  & {\tt ACAT$\$_2$} \\
          5  & {\tt $\$_3$}        & 2 & 7  & 0 & {\tt ACTCAT$\$_1$} \\
          6  & {\tt C}             & 1 & 3 & 1  & {\tt AG$\$_3$} \\
          7  & {\tt C}             & 1 & 3  & 0 & {\tt AT$\$_1$} \\
          8  & {\tt C}             & 2 & 3 & 0  & {\tt AT$\$_2$} \\
          9  & {\tt T}             & 0 & 6  & 0 & {\tt CACAT$\$_2$} \\
          10 & {\tt T}             & 2 & 4 & 1  & {\tt CAG$\$_3$} \\
          11 & {\tt A}             & 2 & 4  & 0 & {\tt CAT$\$_1$}  \\
          12 & {\tt T}            & 3 & 4 & 0   & {\tt CAT$\$_2$} \\
          13 & {\tt G}             & 1 & 6 & 1  & {\tt CTCAG$\$_3$} \\
          14 & {\tt A}             & 4 & 6  & 0 & {\tt CTCAT$\$_1$} \\
          15 & {\tt A}             & 0 & 2 & 1  & {\tt G$\$_3$}  \\
          16 & {\tt $\$_2$}        & 1 & 7 & 1  & {\tt GCTCAG$\$_3$} \\
          17 & {\tt A}             & 0 & 2  & 0 & {\tt T$\$_1$}  \\
          18 & {\tt A}             & 1 & 2 & 0  & {\tt T$\$_2$} \\
          19 & {\tt $\$_1$}        & 1 & 7  & 0 & {\tt TCACAT$\$_2$}  \\
          20 & {\tt C}             & 3 & 5 & 1  & {\tt TCAG$\$_3$} \\
          21 & {\tt C}             & 3 & 5  & 0 & {\tt TCAT$\$_1$}\\
        \cline{2-6}
        \end{tabular}%
    \caption{(c)} \label{fig:mergeEGap}
  \end{subfigure}
  \end{minipage}
}
\caption{The \BWT, \LCP and \SL arrays output by eGap for genomes
    (a) $\S_1$ and (b) $\S_2$. (c) Merged \BWT, \LCP, \SL arrays and
    \DA for $\S_1\S_2$. The context column is not produced by eGap.}
    \label{fig:s1s2eGap}

\end{figure}

\paragraph{Phase 2:}
For each pair of genomes $\S_i$ and $\S_j$,
\gcBB constructs the colored BOSS representation for $\S_i$ and
$\S_j$ from the merged $\BWT_{i,j}$ and $\LCP_{i,j}$ array
as described in~\cite{egidi2019external}.
The bitvector {\tt colors$[1,m]$} and the array {\tt
  coverage$[1,m]$} are computed, where $m$ is the number of
  edges in the colored de Bruijn graph.

{\gcBB} also computes two extra arrays, $\LCS[1,m]$ and
$\KL[1,m]$.  The \LCS array contains the longest common suffix
between consecutive $k$-mers in \Node and the \KL array contains the
length of each vertex label not including the \$ symbols.  These
arrays are easily obtained from $\LCP_{i,j}$ and $\SL_{i,j}$ arrays.

Consider the merged arrays of genomes $\S_1$ and $\S_2$ obtained in
Phase 1 and $k=3$.  The resulting colored BOSS representation is shown in Figure~\ref{tab:s1s2mergeBOSS}.

\begin{figure}[t]
\begin{center}
\small
\begin{tabular}{c c |r| c c c c |c|c|}
\cline{3-3}\cline{8-9}
  $i$ & $last$ & \Node & $W$ & $W^{-}$ & {\tt colors} & {\tt coverage} & \LCS & \KL \\
\cline{3-3} \cline{8-9}
  1 & 1 & {\tt $\$_1$}    & {\tt T}     & 1  & 0 & 2 & 0 & \red{0}  \\
  2 & 1 & {\tt $\$_3$}    & {\tt G}     & 1  & 1 & 1 & 0 & \red{0}  \\
  3 & 1 & {\tt ACA}       & {\tt C}     & 1  & 0 & 1 & 0 & 3   \\
  4 & 1 & {\tt TCA}       & {\tt $\$_3$}& 1  & 0 & 1 & 2 & 3   \\
  5 & 1 & {\tt $\$_3$GA}  & {\tt C}     & 1  & 1 & 1 & 1 & \red{2} \\
  6 & 1 & {\tt $\$_1$TA}  & {\tt C}     & 1  & 0 & 2 & 1 & \red{2}  \\
  8 & 1 & {\tt CAC}       & {\tt T}     & 1  & 0 & 1 & 0 & 3 \\
  9 & 1 & {\tt GAC}       & {\tt T}     & 0  & 1 & 1 & 2 & 3 \\
  10 & 0 & {\tt TAC}       & {\tt A}     & 1  & 0 & 1 & 2 & 3\\
  11 & 1 & {\tt TAC}       & {\tt T}     & 0  & 0 & 1 & 3 & 3 \\
  12 & 0 & {\tt CTC}       & {\tt A}     & 1  & 0 & 1 & 1 & 3  \\
  13 & 1 & {\tt CTC}       & {\tt G}     & 1  & 1 & 1 & 4 & 3  \\
  14 & 1 & {\tt $\$_3$G}   & {\tt A}     & 1  & 1 & 1 & 0 & \red{1} \\
  15 & 1 & {\tt TCG}       & {\tt $\$_2$}& 1  & 1 & 1 & 1 & 3   \\
  16 & 1 & {\tt $\$_1$T}   & {\tt A}     & 1  & 0 & 1 & 0 & \red{1} \\
  17 & 0 & {\tt ACT}       & {\tt $\$_1$}& 1  & 0 & 1 & 1 & 3  \\
  18 & 0 & {\tt ACT}       & {\tt C}     & 1  & 0 & 1 & 3 & 3  \\
  19 & 1 & {\tt ACT}       & {\tt C}     & 0  & 1 & 1 & 3 & 3  \\
\cline{3-3} \cline{8-9}
\end{tabular}%
\end{center}
\caption{$\S_1\S_2$ merged colored BOSS representation with
  $k=3$. Lines where \KL values are colored red represent edges that will be filtered out in the BWSD computation.
}
\label{tab:s1s2mergeBOSS}
\end{figure}

\paragraph{Phase 3:}

The distances between each pair of genomes
$\S_i$ and $\S_j$ are computed by evaluating
the BWSD on the {\tt colors} bitvector, obtaining the corresponding
entry in the expectation and entropy distance matrices.

Note that the colored BOSS representation contains the edges of
every $k'$-mer from the merged genomes, for $1 \leq k' \leq k$.
These edges are part of the BOSS representation and are needed by
the navigation operations~(see \cite{wabi/BoweOSS12}).
%
%Nevertheless, we considered that these $i$-mers where $i < k+1$ can
%compromise our comparisons, since comparing a pair of genomes using
%a value of $(k+1)$ and comparing this same pair using a value
%$(k'+1) > (k+1)$, all $i$-mers, where $1 \leq i \leq k+1$, will be
%considered in both representations.}  Since we are just interested
%in the $(k+1)$-mers for the comparisons, we filtered out all the
%edges of the colored BOSS representation using the \SL array during
%the BWSD computation.  That is, for every edge in the
%representation, we only considered edges where $\SL[i] \geq k+1$.
%
Since we are interested only in the $k$-mers for the comparisons, we
filtered out all the edges of the colored BOSS where $\KL[j] < k$,
for $1 \leq j \leq m$, during the BWSD computation.

%Filtering these edges we can compute the BWSD over the {\tt colors} bitvector, which will represent a comparison on the $(k+1)$-mers of the graph.

From the colored BOSS shown in Figure~\ref{tab:s1s2mergeBOSS},
filtering edges representing $k'$-mers of size smaller than $k$
and using the bitvector {\tt colors} as the bitvector $\alpha$ of the BWSD, we have
$    \alpha = \{0,0,0,1,0,0,0,1,1,0,0,1\},
    r = 0^3 1^1 0^3 1^2 0^2 1^1,
    t_1 = 2, t_2=2, t_3=2 \text{ and } s=6
$
Hence, the BWSD($\S_1,\S_2$) is
\[
P\{k_j=1\}=\frac{2}{6}, P\{k_j=2\}=\frac{2}{6},
P\{k_j=3\}=\frac{2}{6}
\]
Computing the distances we have
$D_M(\S_1,\S_2)=1$ and $D_E(\S_1,\S_2)=1.584$.

\paragraph{Coverage.}

We can use the coverage information in the BWSD to weight the edges of
the graph, aiming at improving the accuracy of the
distance measures.

The same $(k+1)$-mer from distinct genomes can be detected in the
colored BOSS using the \LCS array and the {\tt colors} bitvector.
%
%as follows.  Whenever $\LCP[j] \geq k$, $\LCP[j+1] \geq k$ and
%${\tt colors}[j] \neq {\tt colors}[j+1]$ we have the same
%$(k+1)$-mer from distinct genomes.
%
These repeated $(k+1)$-mers will appear in the $\alpha$ array
with a 0 followed by a 1.  Note that this happens only once
independently of the number of times these $(k+1)$-mers occurred in
both genomes.  Whenever these repeated $(k+1)$-mers occurred
many times in both genomes, their distance should
be decreased.

For example, we added the string {\tt ACTC} in sets $\S_1$ and $\S_2$ from  the previous example.
Let ${\S_1}'=\{{\tt
  TACTCA, TACACT, ACTC, ACTC, ACTC}\}$ and ${\S_2}'=\{{\tt GACTCG,
  ACTC, ACTC}\}$. In both genomes we have to increment the coverage
information of the $k$-mers {\tt ACT} with the outgoing edge {\tt
  C}. The updated lines of the BOSS representation are shown in
Figure~\ref{fig:s1s2mergeBOSScoverage}.

\begin{figure}[t]
\begin{center}
\small
\begin{tabular}{c r |r| c c c c |c|c|}
\cline{3-3}\cline{8-9}
  $i$ & $last$ & \Node & $W$ & $W^{-}$ & {\tt color} & {\tt coverage} & \LCS & \KL  \\
\cline{3-3} \cline{8-9}
  $\vdots$ & $\vdots$ & $\vdots$ & $\vdots$ & $\vdots$ & $\vdots$ & $\vdots$ & $\vdots$ & $\vdots$ \\
  18 & 0 & {\tt ACT}       & {\tt C}     & 1  & 0 & 4 & 3 & 3 \\
  19 & 1 & {\tt ACT}       & {\tt C}     & 0  & 1 & 3 & 3 & 3 \\
\cline{3-3} \cline{8-9}
\end{tabular}%
\end{center}
\caption{Lines with {\tt coverage} incremented for the $(k+1)$-mer {\tt ACTC} in the colored BOSS for $\{{\S_1}',{\S_2}'\}$.}
\label{fig:s1s2mergeBOSScoverage}
\end{figure}

% \begin{figure}[t]
% \begin{center}
% \small
% \begin{tabular}{c r |r| c c c c |c|c|l|}
% \cline{3-3}\cline{8-10}
%   $i$ & $last$ & \Node & $W$ & $W^{-}$ & {\tt color} & {\tt coverage} & \LCP & \SL & context \\
% \cline{3-3} \cline{8-10}
%   $\vdots$ & $\vdots$ & $\vdots$ & $\vdots$ & $\vdots$ & $\vdots$ & $\vdots$ & $\vdots$ & $\vdots$ & $\vdots$\\
%   18 & 0 & {\tt ACT}       & {\tt C}     & 1  & 0 & 4 & 3 & 5 & {\tt TCAG$\$_3$} \\
%   19 & 1 & {\tt ACT}       & {\tt C}     & 0  & 1 & 3 & 3 & 5 & {\tt TCAT$\$_1$} \\
% \cline{3-3} \cline{8-10}
% \end{tabular}%
% \end{center}
% \caption{Lines with {\tt coverage} incremented for $(k+1)$-mer {\tt ACTC} in ${\S_1}^c{\S_2}^c$ BOSS.}
% \label{fig:s1s2mergeBOSScoverage}
% \end{figure}

Let $\alpha' =
\{0,0,0,1,0,0,0,1,1,0,0,1\}$ be a bitvector equal to $\alpha$ from the previous example.
%\orange{Não entendi o que é esse take.  Está só renomeando? É o mesmo ou mudou?}.
The last 0 and 1 values from $\alpha'$
represent the $(k+1)$-mer {\tt ACTC} from both genomes. We apply the coverage value to the positions of $r'$ where
these values occurred. That is, $r'  = 0^3 1^1 0^3 1^2 \blue{0^{1+3} 1^{1+2}} = 0^3 1^1 0^3 1^2 \blue{0^4 1^3}$. Finally, we expand $r'$
in the positions of the equal $(k+1)$-mers while merging them, that is
$r' = 0^3 1^1 0^3 1^2 0^1 \blue{0^1 1^1 0^1 1^1 0^1 1^1 0^1 1^0}$.
Then, we have $t_1 =
8$, $t_2=1$, $t_3=2$ and $s=11$.
And the BWSD(${\S_1}',{\S_2}'$)
is
\[P\{k_j=1\}=\frac{9}{12}, P\{k_j=2\}=\frac{1}{12},
P\{k_j=3\}=\frac{2}{12}\]

\noindent
Computing the distances we have
$D_M({\S_1}'{\S_2}')=0.41666$ and $D_E({\S_1}'{\S_2}')=1.04085$.
The effect of coverage on the similarity is analysed in our experiments.
%\green{Therefore we can observe that the use of coverage information can modify the BWSD computed distances.
%This difference will be explored during the experiments.}

\paragraph{Time and space analysis.}

Let $N_1$ and $N_2$ be the sizes of two genomes.
Phase 1 takes $O((N_1+N_2) \maxlcp)$ time to construct and merge the
\BWT, \LCP and \SL in external memory with eGap, where $\maxlcp$ is
the maximum in LCP.

Phase 2 takes $O(N_1+N_2)$ time to construct the BOSS representation.
%In the BOSS construction during phase 2 we have a sequential scan over \LCP, \BWT and \DA, which takes $O(N)$ time.

Let $m$ be the number of edges in the colored BOSS.
The space required for the BOSS
representation is $5m+6\log_2 m$ bits, as shown in
Section~\ref{sec:deBruijnBOSS}.
The {\tt colors} bitvector and the
{\tt coverage} array require extra $m$ bits and $4m$ bytes
respectively. For reads with less than 65K symbols both \LCS and \KL can be stored
in arrays of short integers, that is, $2m$ bytes for each one.
Therefore, the overall space required is $8m$ bytes plus $6m+6\log_2 m$ bits.

Phase 3 takes $O(m)$ time to compute the BWSD from the  {\tt colors} bitvector, \LCS and \KL arrays.
%In the BWSD computation during phase 3 we have a sequential scan over {\tt colors} bitvector and \SL array, this takes $O(m)$ time.
The arrays $r$ and $t$ require $O(m)$ bytes.  %The running time for a pair of genomes is $O(N \maxlcp)+ O(N) + O(m)$ and the required space is $O(m)$.

\subsection{An improved algorithm: \mgcBB}
\label{s:mgcbb}

In this section, we describe an extension to \gcBB, called \mgcBB,
that computes the colored BOSS for all genomes $\S_1,\S_2,\ldots,
\S_g$ only once, instead of constructing pairwise as above,
and computes the distance matrices $\DM$ and $\DE$ using
  compressed data structures, as proposed in \cite{tcs/LouzaTGZ19}
for the BWSD.
The pseudo-code for \mgcBB is shown in Algorithm~\ref{alg:mgcBB}.

In Phase 1, $\mgcBB$ constructs the $\BWT$ and the arrays $\LCP$ and
CL for each genome $\S_1, \S_2, \dots, \S_g$ with eGap (as in the
previous version).  Then, all these arrays are merged only once, computing the document array
$\DA$ for all genomes as well.

In Phase 2, $\mgcBB$ computes the colored BOSS for all genomes, modifying {\tt colors$[1,m]$} to be an
integer array and $m$  the number of edges in the colored de Bruijn graph for $\S_1, \S_2, \dots, \S_g$.

In Phase 3, \mgcBB builds $g$ bitvectors $B_i$, with $|B_i|=m$,
where $B_i[j]=1$ if {\tt colors$[j] = i$} or $B_i[j]=0$ otherwise.
For each bitvector $B_i$, an $O(1)$ rank/select
structure~\cite{Munro96} is built.  The algorithm then proceeds line
by line on the matrix.  To evaluate the distances among $S_i$ and
$S_{j>i}$, the algorithm selects the intervals over {\tt colors}
that contain consecutive occurrences of $i$.  For each interval the
algorithm counts the $k_j$ occurrences of $j$, which corresponds to
the existence of the run $1^{k_j}$ in the sequence of runs for $S_i$
and $S_j$.  The runs $0^{\ell_j+1}$ are computed when $\ell_j$
consecutive intervals do not contain any occurrence of $j$.

\begin{algorithm}[!ht]
\SetAlgoLined
\KwIn{Collection $\S$ and selected $k$}
\KwOut{Matrices $D_M$ and $D_E$ of double precision numbers}
\tcp{Phase 1}
\For{each genome $\S_i$ in \S}{
    ${\tt eGap}(\S_i)$\tcp*{compute \LCP, \BWT and \SL}
}
${\tt eGap\_merge(\S)}$\tcp*{merge \LCP, \BWT and \SL and generate \DA}
$\text{double } D_M[1..d][1..d]=0.0$\;
$\text{double } D_E[1..d][1..d]=0.0$\;
\tcp{Phase 2}
${\tt colored\_boss\_construction}(\LCP, \BWT, \SL, \DA, k)$\tcp*{compute {\tt colors}}
\tcp{Phase 3}
${\tt bwsd\_all}({\tt colors}, \LCP, \SL, D_M, D_E)$\;
\Return{$D_M, D_E$}\;
\caption{mgcBB}
\label{alg:mgcBB}
\end{algorithm}

\paragraph{Time and space analysis.} % (mgcBB)}
%To construct and merge the \BWT, \LCP and \SL in Phase 1 with eGap takes $O(N \maxlcp)$ time, where $\maxlcp$ is the maximum in LCP.
Let $\S$ be a collection of $g$ genomes and $N$ be the total length of all genomes.
Phase 1 takes time $O(N \maxlcp)$ time to construct and merges all arrays with eGap.
%, where $\maxlcp$ is the maximum in $\LCP$.
Phase 2 takes $O(N)$ time to construct the BOSS representation.
%Let $m$ be the number of edges in the colored BOSS.
%The space required for the BOSS
%representation is $5m+6\log_2 m+m\log$ bits, as shown in
%Section~\ref{sec:deBruijnBOSS}.
The {\tt colors} array now requires $m\log_2 g$ bits.

Then, the overall space required for the BOSS
representation with \LCS and \KL arrays is $5m + 6\log_2 m + m \log_2 g$ bits plus $8m$ bytes.

%and the
%{\tt coverage} array require extra $4m$ bytes each.
%For reads with less than 65K symbols both \LCS and \KL can be stored in arrays of short integers, that is, $2m$ bytes for each one.
%Therefore, the overall space required is $6m$ bytes plus $6m+6\log_2 m$ bits.

Phase 3 takes $O(dN)$ time to compute the BWSD from the {\tt colors} array, \LCS and \KL arrays.
The bitvectors $B_i$ with support to rank/select queries require $dN + o(dN)$ bits.
The arrays $r$ and $t$ require $O(N)$ bytes.

\section{Experiments}

We evaluated \gcBB by reconstructing the phylogeny of the 12
Drosophila species in Table~\ref{tab:drosophilasGenomesInfo},
obtained from FlyBase~\cite{flybase}.  The reads were obtained with a NextSeq 500 sequencer\footnote{\url{https://www.illumina.com/systems/sequencing-platforms/nextseq.html}} and have 302 bp on average, except that reads of {\it D.~grimshawi}
were obtained with a MinION
sequencer\footnote{\url{https://nanoporetech.com/products/minion}} and have 6,520 bp on the average.
The phylogenies in the sequel were drawn using iTOL~\cite{itol}.

\begin{table}[!ht]
    \centering
    \caption{Information on the genomes of Drosophilas, that can be
      accessed through their Run (SRR) or BioSample (SAMN)
      accessions at \url{https://www.ncbi.nlm.nih.gov/genbank/}.
      The Bases column has the number of sequenced bases in Gbp. The
      Reference column has the size of the complete genome in Mb.}
    \label{tab:drosophilasGenomesInfo}
    \begin{tabular}{|l|c|c|r|r|}
        \hline
        Organism & Run & BioSample & Bases (Gbp) & Reference (Mb)  \\
        \hline
        {\it D. melanogaster} & SRR6702604 & SAMN08511563 & 6.20 & 138.93 \\
        {\it D. ananassae} & SRR6425991 & SAMN08272423 & 7.13 & 215.47 \\
        {\it D. simulans} & SRR6425999 & SAMN08272428 & 9.22 & 131.66 \\
        {\it D. virilis} & SRR6426000 & SAMN08272429 & 11.16 & 189.44 \\
        {\it D. willistoni} & SRR6426003 & SAMN08272432 & 11.66 & 246.98  \\
        {\it D. pseudoobscura} & SRR6426001 & SAMN08272435 & 12.28 & 163.29 \\
        {\it D. mojavensis} & SRR6425997 & SAMN08272426 & 12.45 & 163.17 \\
        {\it D. yakuba} & SRR6426004 & SAMN08272438 & 12.78 & 147.90 \\
        {\it D. persimilis} & SRR6425998 & SAMN08272433 & 13.32 & 195.51 \\
        {\it D. erecta} & SRR6425990 & SAMN08272424 & 14.01 & 146.54  \\
        {\it D. sechellia} & SRR6426002 & SAMN08272427 & 14.44 & 154.19 \\
        {\it D. grimshawi} & SRR13070661 & SAMN16729613 & 14.50 & 191.38 \\
        \hline
    \end{tabular}
\end{table}

\begin{figure}
    \centering
    \includegraphics[width=0.9\textwidth]{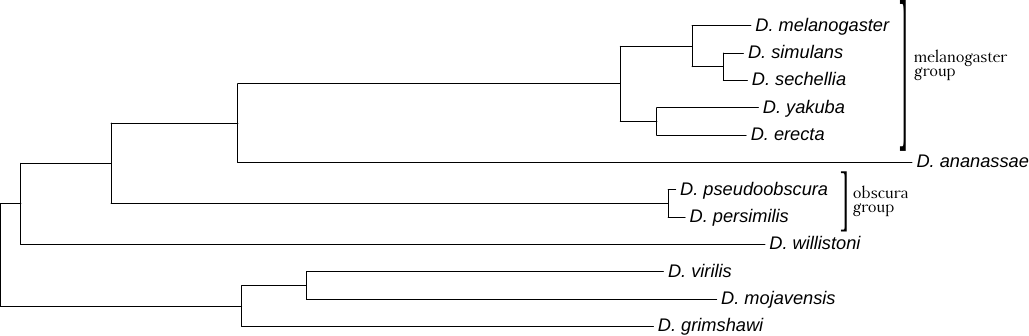}
    \caption{Drosophila phylogeny, after~\cite{clark2007evolution}.}
    \label{fig:drosophila_subgroups}
\end{figure}

Figure~{\ref{fig:drosophila_subgroups}} shows a phylogeny for the 12
Drosophila genomes~{\cite{clark2007evolution,hahn2007gene}} built
using Neighbor-Joining on distances inferred on alignments among
families of genes.  This phylogeny was as reference, that
is, we analysed if the distances computed by \gcBB lead to a
phylogeny that agrees with this phylogeny.

Our algorithms were implemented in C and compiled
with {\tt gcc} version 4.9.2.  Our prototype implementation can be
accessed at \url{https://github.com/lucaspr98/gcBB}.  As previously mentioned, we used eGap~\cite{egidi2019external} to construct and merge the data structures during Phase 1.  The
experiments were conducted on a system with Debian GNU/Linux 4.9.2
64 bits on Intel Xeon E5-2630 v3 20M Cache 2.40 GHz processors, 378
GB of RAM and 13 TB SAS storage. Our experiments were limited to 48 GB of RAM.

\subsection{Running time}

The times to run eGap to compute the BWT and the LCP and {\SL} arrays for each genome $S_i \in \S$ in Phase 1 are shown in Table~{\ref{tab:genomesStructuresInfo}}. The longest running time was approximately 57 hours, with the resulting arrays taking about 68 GB of space on disk.  The sizes of the BWT, the sizes of the LCP and {\SL} arrays, and the average and maximum LCP values are also shown in Table~{\ref{tab:genomesStructuresInfo}}.  Both LCP and {\SL} arrays use $2$ bytes per entry.

\begin{table}[t]
    \centering
    \caption{Construction information on data structures for the Drosophilas genomes.}
    \begin{tabular}{|l|r|r|r|r|r|r|}
        \hline
        Organism & BWT & LCP & CL & Running time & LCP avg & LCP max  \\
        \hline
        {\it D. melanogaster} & 5.9 GB & 12 GB & 12 GB &  17.30h & 61.54 & 302\\
        {\it D. ananassae}    & 6.7 GB & 14 GB & 14 GB & 18.37h & 55.56 & 302 \\
        {\it D. simulans}     & 8.7 GB & 18 GB & 18 GB & 24.18h & 58.96 & 302\\
        {\it D. virilis}      & 11 GB  & 21 GB & 21 GB & 28.45h & 55.06 & 302 \\
        {\it D. willistoni}   & 11 GB  & 22 GB & 22 GB & 37.31h & 57.79 & 302 \\
        {\it D. pseudoobscura}& 12 GB  & 23 GB & 23 GB & 40.07h & 58.72 & 302\\
        {\it D. mojavensis}   & 12 GB  & 24 GB & 24 GB &  32.65h & 58.09 & 302\\
        {\it D. yakuba}       & 12 GB  & 24 GB & 24 GB &  42.42h & 59.98 & 302\\
        {\it D. persimilis}   & 13 GB  & 25 GB & 25 GB & 35.30h & 58.76 & 302\\
        {\it D. erecta}       &  14 GB & 27 GB & 27 GB &  42.92h & 60.47 & 302\\
        {\it D. sechellia}    &  14 GB & 27 GB & 27 GB &  37.88h & 61.25 & 302\\
        {\it D. grimshawi}    &  14 GB & 27 GB & 27 GB &  57.40h & 43.07 & 2648 \\
        \hline
    \end{tabular}
    \label{tab:genomesStructuresInfo}
\end{table}

The overall time to run eGap to merge the computed data structures,
for all pairs of genomes $\{S_i,S_j\} \in \S$, with the pairwise
approach (Algorithm~{\ref{alg:gcBB}}) was approximately 154 days,
whereas the overall time to run eGap to merge the data structures of
all genomes in $\S$ at once with the all-vs-all approach
(Algorithm~{\ref{alg:mgcBB}}) was approximately 20 days.
The size of the merged files was approximately the sum of the sizes
of the input files.  The document array file has the same size of
the merged {\BWT} file, since both store each value using one byte.

In Phase 2, the overall running time to build all pairs of colored
BOSS data structures with the pairwise approach was approximately 57
hours, while the average time to build the colored BOSS for a pair $\{S_i,S_j\}$ was about 52
minutes.  The time to build the colored BOSS for all
genomes at once with the all-vs-all approach was approximately 5.5
hours.

In Phase 3, the overall running time to compute all pairs of BWSD
distance matrices entries with the pairwise approach was
approximately 1.7 hours.  The average time was about 2 minutes
for each pair.  The time to compute the BWSD distance matrices at
once with the all-vs-all approach was approximately 10.8 hours.

Therefore, we spent approximately 156.4 days with \gcBB
against 20.6 days using \mgcBB. The running times for Phases 1 and 2 were obtained with coverage information and $k=31$.

%\todo{disk space?}

\subsection{Phylogenetic trees}

We ran \gcBB for $k=15$, $31$ and $63$, producing entropy and
expectation BWSD distance matrices for the 12 Drosophilas, with and without coverage information.

We used the Neighbor-Joining~\cite{saitouNeighborJoining} on the distance
  matrices to reconstruct the phylogenies. Following the
construction for the reference phylogeny, we executed
Neighbor-Joining downto $n=2$ to root the phylogeny.
We used the
Robinson-Foulds~\cite{robinson1981comparison} distance to compare
our phylogenies with the reference phylogeny.

Let $T$ be a phylogenetic tree with $n$ vertices labeled by 
$U = \{1, 2, \ldots , n\}$. If an edge is removed from $T$ then it induces 
a bipartition of $U$.  If every edge in the set $E$ of edges of $T$ is removed in turn, a set of induced bipartitions $T(E)$ is defined.
For a pair of phylogenetic trees $T_1$ and $T_2$ with the same set of labelled leaves, the Robinson-Foulds distance between $T_1$ and $T_2$ is the size of the symmetric difference of $T_1(E_1)$ and $T_2(E_2)$; it is a metric whose values vary from $0$ to $2n-6$.

\begin{figure}[!ht]
  \centering
  \begin{subfigure}[t]{0.5\textwidth}
    \includegraphics[width=\linewidth]{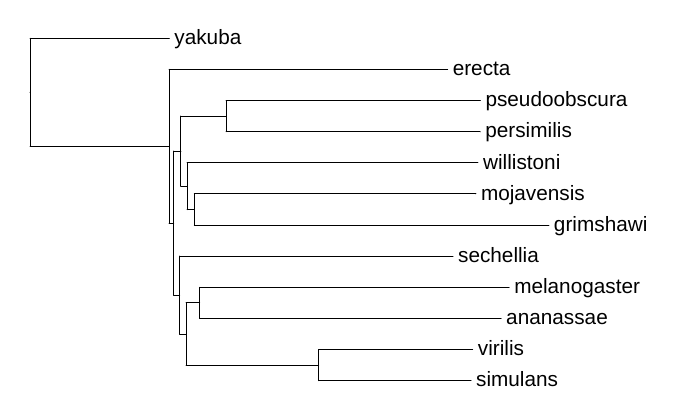}
    \caption{(a)} \label{fig:drosophilas_final_16_entropy}
  \end{subfigure}%
  \hspace*{\fill}
  \begin{subfigure}[t]{0.5\textwidth}
    \includegraphics[width=\linewidth]{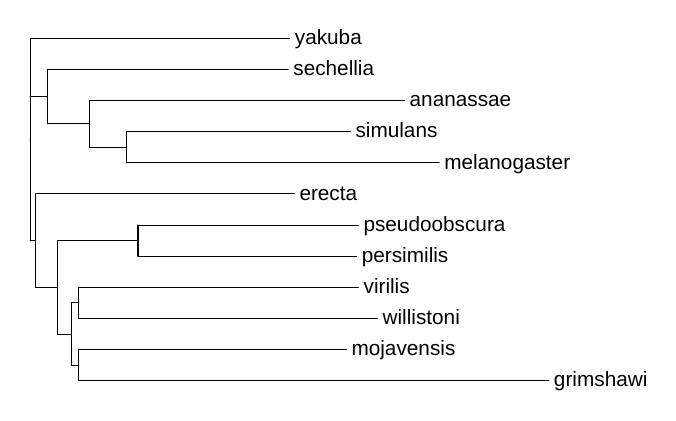}
    \caption{(b)} \label{fig:drosophilas_final_16_expectation}
  \end{subfigure}%
  \caption{\gcBB phylogenies with $k=15$, (a) using entropy, (b) using expectation.}
    \label{fig:drosophilas_final_16}
  \begin{subfigure}[t]{0.5\textwidth}
    \includegraphics[width=\linewidth]{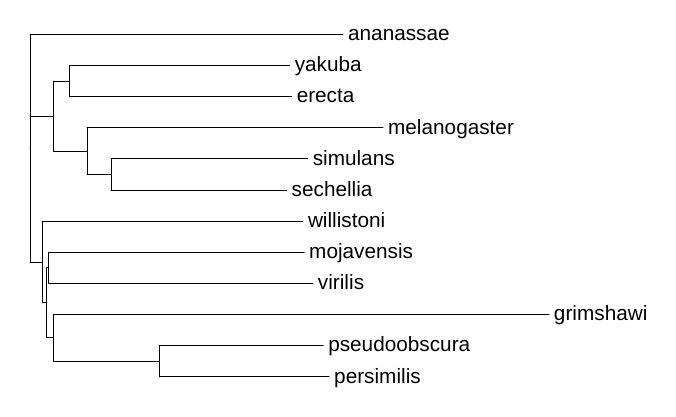}
    \caption{(a)} \label{fig:drosophilas_final_16_entropy_coverage}
  \end{subfigure}%
  \hspace*{\fill}
  \begin{subfigure}[t]{0.5\textwidth}
    \includegraphics[width=\linewidth]{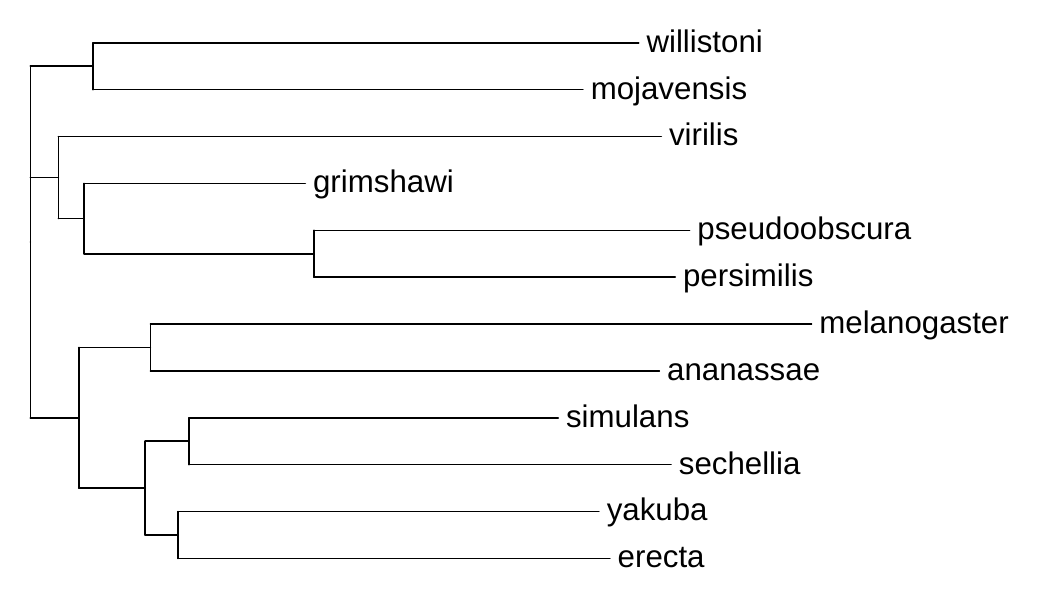}
    \caption{(b)} \label{fig:drosophilas_final_16_expectation_coverage}
  \end{subfigure}%
  \caption{\gcBB phylogenies with $k=15$ and coverage information, (a) using entropy, (b) using expectation.}
  \label{fig:drosophilas_final_16_coverage}
\end{figure}

%By including coverage information in the BWSD computation, we obtained the phylogenies shown in

Figures \ref{fig:drosophilas_final_16} and
\ref{fig:drosophilas_final_16_coverage} show the phylogenies for $k=15$
with and without coverage information, respectively.  The pair {\it
  D. pseudoobscura} and {\it D. persimilis} from the {\tt obscura}
group agrees with the reference phylogeny.  With coverage
information there is a clear separation between the
\texttt{melanogaster} group and the outer groups.  However, in
general, the placement of the other genomes disagree with the
reference phylogeny.

\begin{figure}[!ht]
  \centering
  \begin{subfigure}[t]{0.5\textwidth}
    \includegraphics[width=\linewidth]{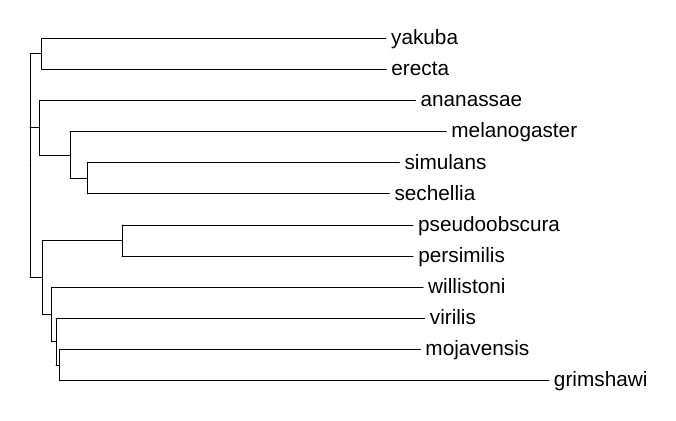}
    \caption{(a)} \label{fig:drosophilas_final_32_entropy}
  \end{subfigure}%
  \hspace*{\fill}
  \begin{subfigure}[t]{0.5\textwidth}
    \includegraphics[width=\linewidth]{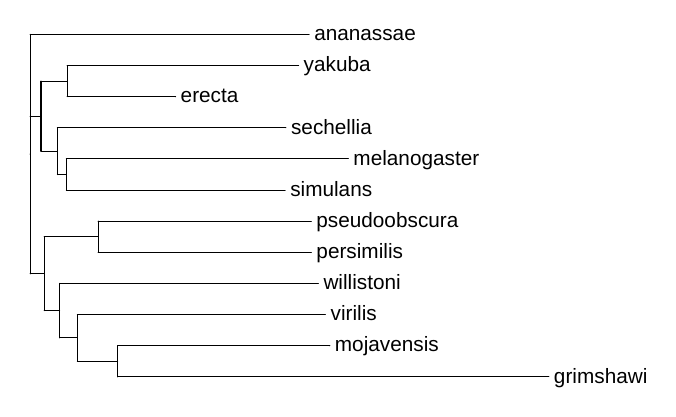}
    \caption{(b)} \label{fig:drosophilas_final_32_expectation}
  \end{subfigure}%
  \caption{\gcBB phylogenies with $k=31$, (a) using entropy, (b) using expectation.}
    \label{fig:drosophilas_final_32}
  \begin{subfigure}[t]{0.5\textwidth}
    \includegraphics[width=\linewidth]{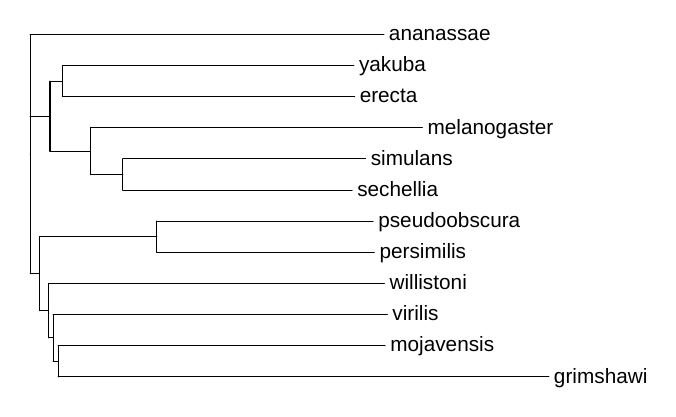}
    \caption{(a)} \label{fig:drosophilas_final_32_entropy_coverage}
  \end{subfigure}%
  \hspace*{\fill}
  \begin{subfigure}[t]{0.5\textwidth}
    \includegraphics[width=\linewidth]{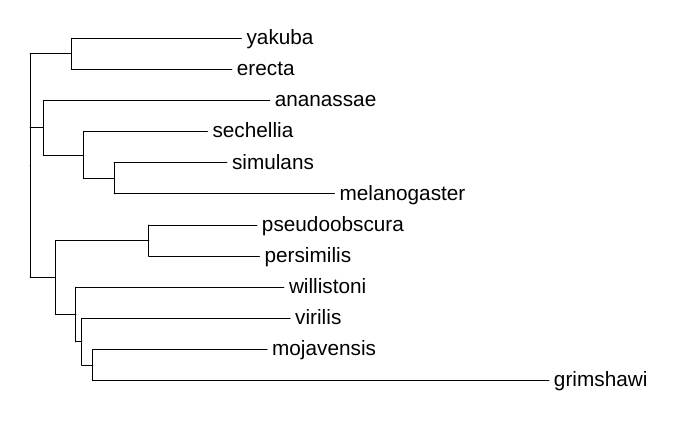}
    \caption{(b)} \label{fig:drosophilas_final_32_expectation_coverage}
  \end{subfigure}%
  \caption{\gcBB phylogenies with $k=31$ and coverage information, (a) using entropy, (b) using expectation.}
    \label{fig:drosophilas_final_32_coverage}
\end{figure}

Figures~{\ref{fig:drosophilas_final_32}} and
{\ref{fig:drosophilas_final_32_coverage}} show the phylogenies for
$k=31$ with and without coverage information, respectively.
There is one inconsistency
involving \textit{D. grimshawi} and \textit{D. virilis}, which were
swapped in our phylogenies, but are in the same subtree.
Nonetheless, the high level groups division agrees with the
reference phylogeny.

%The phylogenies are basically the same.

%Again, there is a division between the \texttt{melanogaster} ad \texttt{obscura} groups with the remaining groups together in a subtree. We can observe a few inconsistencies inside the subgroups, but the high level groups division agrees with the reference phylogeny.

%\orange{Quase nunca há correct em filogenia, uma filogenia é sempre uma hipótese.}

\begin{figure}[!ht]
  \centering
  \begin{subfigure}[t]{0.5\textwidth}
    \includegraphics[width=\linewidth]{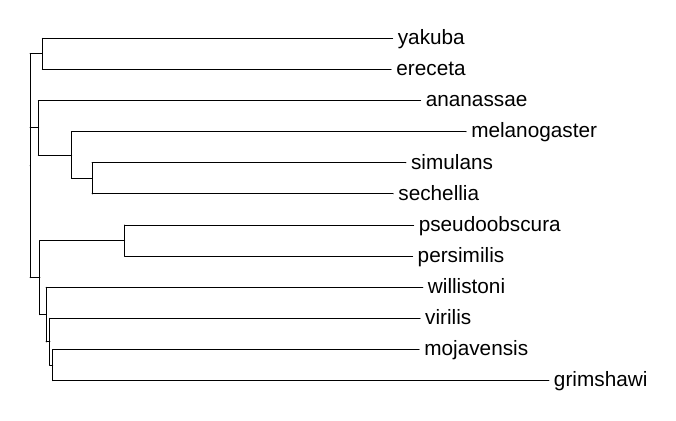}
    \caption{(a)} \label{fig:drosophilas_final_64_entropy}
  \end{subfigure}%
  \hspace*{\fill}
  \begin{subfigure}[t]{0.5\textwidth}
\includegraphics[width=\linewidth]{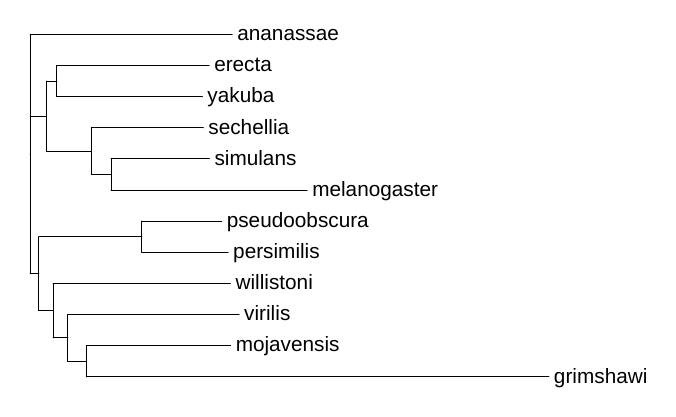}
    \caption{(b)} \label{fig:drosophilas_final_64_expectation}
  \end{subfigure}%
  \caption{\gcBB phylogenies with $k=63$, (a) using entropy, (b) using expectation.}
    \label{fig:drosophilas_final_64}
  \begin{subfigure}[t]{0.5\textwidth}
    \includegraphics[width=\linewidth]{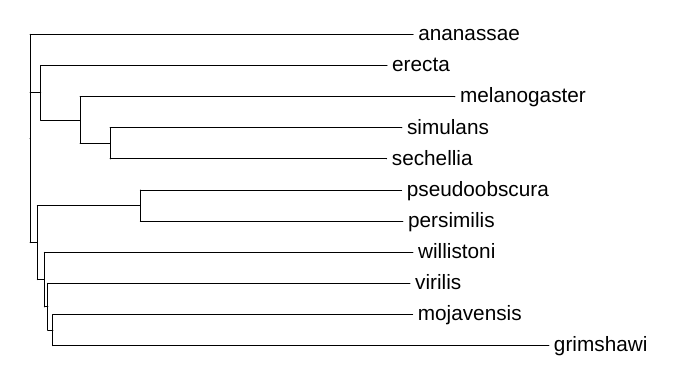}
    \caption{(a)} \label{fig:drosophilas_final_64_entropy_coverage}
  \end{subfigure}%
  \hspace*{\fill}
  \begin{subfigure}[t]{0.5\textwidth}
     \includegraphics[width=\linewidth]{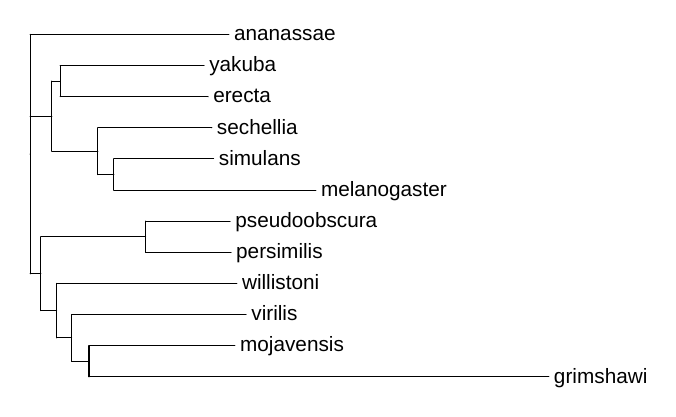}
    \caption{(b)} \label{fig:drosophilas_final_64_expectation_coverage}
  \end{subfigure}%
  \caption{\gcBB phylogenies with $k=63$ and coverage information, (a) using entropy, (b) using expectation.}
    \label{fig:drosophilas_final_64_coverage}
\end{figure}

Figures~{\ref{fig:drosophilas_final_64}} and
{\ref{fig:drosophilas_final_64_coverage}} show the phylogenies for
$k=63$ with and without coverage information, respectively.  They
are very similar to those resulting with $k=31$.

Table~\ref{tab:robinsonFinal} shows the Robinson-Foulds distance
evaluated between the phylogenies by \gcBB and the reference
phylogeny.  The phylogenies by {\gcBB} which are closer to the
reference were constructed using $k=31$ and $k=63$, with coverage
information and entropy distance.

\begin{table}
    \centering
    \caption{Robinson-Foulds distances computed between phylogenies by \gcBB and the reference phylogeny of Drosophila genomes in Table~\ref{tab:drosophilasGenomesInfo}. The symbol c indicates the phylogenies constructed by \gcBB using coverage information.}
    \begin{tabular}{l|c|c|c|c|c|c|}
        \cline{2-7}
        & 15 & 15c & 31 & 31c & 63 & 63c  \\
        \hline
        \multicolumn{1}{|l|}{Entropy} & 7 & 2 & 2 &  1 &  2 & 1\\

        \hline
        \multicolumn{1}{|l|}{Expectation} & 6 & 5 & 2 & 3  &  2 & 2\\
        \hline
    \end{tabular}
    \label{tab:robinsonFinal}
\end{table}

\paragraph{Effect of data size}
In order to evaluate the effect of read sizes in the resulting
phylogenies, we considered sequencing data from an Illumina HiSeq
2000 for {\it D. grimshawi}. The information on this genome, the
running time taken by eGap to construct the data structures and
their sizes are shown in Table~\ref{tab:smallGrimshawiInfo}.

\begin{table}[!ht]
    \centering
    \caption{Information on the genome of {\it D. grimshawi}, that
      can be accessed through its Run (SRR) or BioSample (SAMN)
      accessions at \url{https://www.ncbi.nlm.nih.gov/genbank/}. The
      Bases column has the number of sequenced bases in Gbp. The
      Reference column has the size of the complete genome in Mb.
      The sizes of data structures in Gb are shown in columns \BWT,
      \LCP and \SL, and the average \LCP is shown in column \LCP
      avg.}

    \begin{tabular}{|l||c|c|r|r|r|r|r|r|}
        \hline
        Organism & Run & BioSample & Bases & Reference & BWT & LCP & \SL & LCP avg  \\
        \hline
        {\it D. grimshawi} & 7642855 & 09764638 & 1.80 & 191.38 &  1.80 & 3.5 & 3.5 & 28.74 \\
        \hline
    \end{tabular}
    \label{tab:smallGrimshawiInfo}
\end{table}

\begin{figure}[t]
  \centering
  \begin{subfigure}[t]{0.5\textwidth}
    \includegraphics[width=\linewidth]{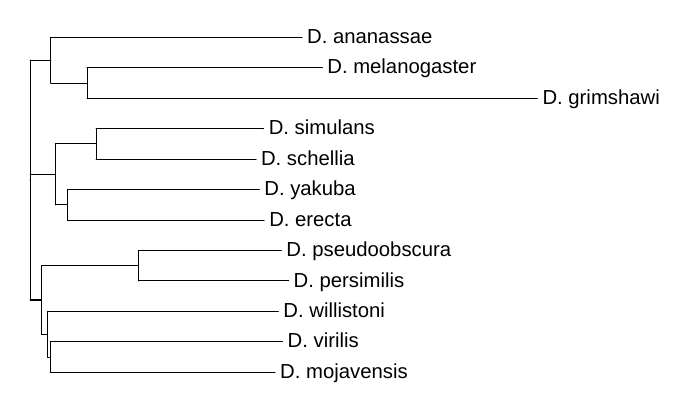}
    \caption{(a)} \label{fig:drosophilas_16_entropy_coverage}
  \end{subfigure}%
  \hspace*{\fill}
  \begin{subfigure}[t]{0.5\textwidth}
    \includegraphics[width=\linewidth]{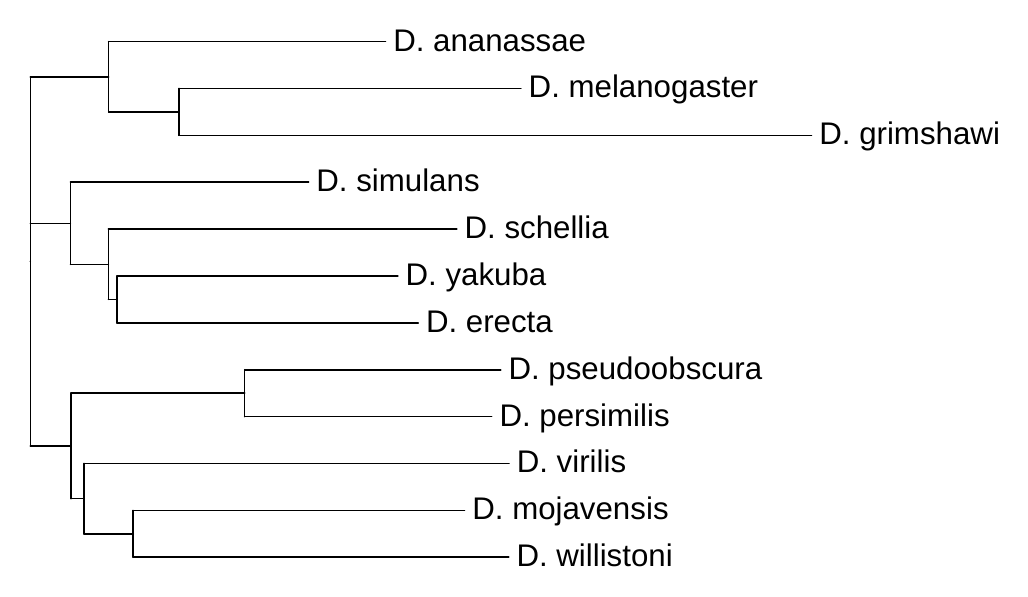}
    \caption{(b)} \label{fig:drosophilas_16_expectation_coverage}
  \end{subfigure}%
  \caption{\gcBB phylogenies for Drosophilas with the alternative {\it D. grimshawi} with $k=15$ and coverage information, (a) using entropy, (b) using expectation.}
    \label{fig:drosophilas_16_coverage}
\end{figure}

We executed \gcBB using the same parameters and values of $k$.  The
best phylogeny was obtained with $k=15$  using coverage
information. It is shown in Figure~{\ref{fig:drosophilas_16_coverage}}.
By computing the Robinson-Foulds distance between these
phylogenies and the reference phylogeny we obtained the values in
Table~\ref{tab:robinsonSmallGrimshawi}.

\begin{table}[t]
    \centering
    \caption{Robinson-Foulds distance computed between phylogenies with {\it D. grimshawi} from another experiment and the reference phylogeny. The symbol c represents the phylogenies constructed using coverage information.}
    \begin{tabular}{l|c|c|c|c|c|c|}
        \cline{2-7}
        & 15 & 15c & 31 & 31c & 63 & 63c  \\
        \hline
        \multicolumn{1}{|l|}{Entropy} & 7 & 5 & 5 &  5 &  5 & 5\\
        \hline
        \multicolumn{1}{|l|}{Expectation} & 6 & 7 & 6 & 5  &  6 & 6\\
        \hline
    \end{tabular}
    \label{tab:robinsonSmallGrimshawi}
\end{table}

We believe that significantly different amounts of sequenced
bases impairs \gcBB in its current form.  In this experiment {\it
  D. grimshawi} has 1.8 Gbp, while {\it D. sechellia} and {\it
  D. simulans} have more than 14 Gbp.
 When constructing the colored de Bruijn graph for {\it
   D. grimshawi} and {\it D. sechellia} there will be much more
 edges from {\it D. sechellia} than from {\it D. grimshawi}, and the
 similarity between these genomes tends to be smaller than it should
 be.  Moreover, when constructing the colored de Bruijn graph for
 {\it D. grimshawi} and {\it D. melanogaster} there will also be
 much more edges from {\it D. melanogaster}.  The difference between
 the amount of bases in {\it D. melanogaster} and {\it D. grimshawi}
 is around 4 Gbp, while from {\it D. sechellia} to {\it
   D. grimshawi} is around 12 Gbp.

These results suggest that our algorithms produce reasonable
phylogenies when $k$ is closer to the average \LCP of the reads. Also,
the coverage information reduced the Robinson-Foulds distance
to the reference in most cases. Finally, the fact that all reads in
the dataset were obtained using similar sequencing protocols and, on
average, have a similar number of sequenced bases may have helped
obtaining favorable results.

\section{Conclusions}

In this work we introduced a new method to compare genomes prior to
assembly using space-efficient data structures implemented as \gcBB
and \mgcBB, algorithms to compare sets of reads of genomes using the
BOSS representation and to compute the similarity measures based on
the BWSD.

We evaluated our algorithms reconstructing the phylogeny of 12 {\it
  Drosophila} genomes.  We used Neighbor-Joining over the distance
matrices output by \gcBB to reconstruct phylogenetic trees.  Then we
computed the Robinson-Foulds distance between the phylogenies by
\gcBB and a reference phylogeny. One issue when working with the de
Bruijn graph is setting the value of $k$. We observed that values
over the average \LCP of the genomes lead to reasonable results. We
observed better results using the entropy measure and coverage
information in the BWSD computation.

Future research may investigate different strategies for dealing
with coverage information, as the experiments indicate a positive
contribution of coverage to the resulting phylogenies.  The quality
of sequenced bases may also be investigated in future work as a
means to improve the method.

\backmatter

% \begin{comment}
%
%
% \bmhead{Supplementary information}
%
% If your article has accompanying supplementary file/s please state so here.
%
% Authors reporting data from electrophoretic gels and blots should supply the full unprocessed scans for key as part of their Supplementary information. This may be requested by the editorial team/s if it is missing.
%
% Please refer to Journal-level guidance for any specific requirements.
% \end{comment}

\paragraph{Acknowledgements.}

%Acknowledgements are not compulsory. Where included they should be brief. Grant or contribution numbers may be acknowledged.
%Please refer to Journal-level guidance for any specific requirements.

%\section*{Acknowledgements}
The authors thank Prof.\ Marinella Sciortino for helpful discussions and thank Prof.\ Nalvo Almeida for granting access to the computer used in the experiments.

\paragraph{Funding.}
L.P.R.\ acknowledges that this study was financed by Coordenação de Aperfeiçoamento de Pessoal de Nível Superior (CAPES), Brazil, Financing Code 001.
F.A.L.\ acknowledges the financial support from CNPq (grants 408314/2023-0 and 406418/2021-7) and FAPEMIG (grant APQ-01217-22).
G.P.T.\ acknowledges the financial support of Brazilian agencies CNPq (grants 408314/2023-0 and 317249/2021-5) and CAPES.

\end{document}